\documentclass[prd,showpacs,showkeys,nofootinbib,floatfix,eqsecnum,fleqn,
               preprint,12pt,tightenlines]{revtex4} 

\usepackage{amsmath,amssymb,revsymb,graphicx,dcolumn}

\newcommand{\beq}{\begin{equation}}
\newcommand{\eeq}{\end{equation}}
\newcommand{\beqa}{\begin{eqnarray}}
\newcommand{\eeqa}{\end{eqnarray}}
\newcommand{\bsubeqs}{\begin{subequations}}
\newcommand{\esubeqs}{\end{subequations}}

\newcommand{\half}{{\textstyle \frac{1}{2}}}    

\newcommand{\dimlessscalar}{s}                  
\newcommand{\dimlesstime}  {t}                  
\newcommand{\dimfultime}   {\tau}               

\begin{document}


\noindent Phys. Rev. D 81, 043006  (2010)
\hfill    arXiv:0904.3276 [gr-qc] 
\vspace*{10mm}
\title{Gluon condensate, modified gravity, and the accelerating Universe\vspace*{5mm}}
\author{F.R. Klinkhamer}
\email{frans.klinkhamer@kit.edu}
\affiliation{\mbox{Institute for Theoretical Physics, University of Karlsruhe,}\\
\mbox{Karlsruhe Institute of Technology, 76128 Karlsruhe, Germany}\\}

\begin{abstract}
\vspace*{.5mm}\noindent
It has been suggested recently to study
the dynamics of a gravitating gluon condensate $q$
in the context of a spatially flat Friedmann--Robertson--Walker universe.
The expansion of the Universe (or, more generally, the presence of a nonvanishing
Ricci curvature scalar $R$) perturbs the gluon condensate and may induce a
nonanalytic term $\widetilde{h}(R,q)$ in the effective gravitational action.
The aim of this article is to explore the cosmological implications of a
particular nonanalytic term $\widetilde{h} \propto \eta\,|R|^{1/2}\,|q|^{3/4}$.
With a quadratic approximation of the gravitating gluon-condensate
vacuum energy density $\rho_{V}(q)$ near the equilibrium value $q_{0}$
and a small coupling constant $\eta$ of the modified-gravity term $\widetilde{h}$,
an ``accelerating universe'' is obtained which resembles
the present Universe, both qualitatively and quantitatively.
The unknown component $X$ of this model universe
(here, primarily due to modified-gravity effects)
has an effective equation-of-state parameter $\overline{w}_{X}$
which is found to evolve toward the value $-1$ from above.
\end{abstract}

\pacs{95.36.+x, 98.80.Jk, 04.20.Cv}
\keywords{dark energy, cosmology, general relativity}
\maketitle

\section{Introduction}\label{sec:introduction}

The fundamental theory of the strong interactions is nowadays taken
to  be quantum chromodynamics (QCD);
see, e.g., Refs.~\cite{ChengLi1985,Marshak1993} and other references therein.
In the framework of this theory, there is evidence for the existence of a gluon
condensate~\cite{ShifmanVainshteinZakharov1978,Narison1996,Rakow2006,AndreevZakharov2007}.
The question, then, is how
the gluon condensate gravitates and evolves as the Universe expands.
Here, a tentative answer is obtained by use of
the so-called $q$--theory approach for the gravitational effects of vacuum energy
density~\cite{KlinkhamerVolovik2008a,KlinkhamerVolovik2008b,KlinkhamerVolovik2008c,KlinkhamerVolovik2009a}.

The outline of this article is as follows.
In Sec.~\ref{sec:Gluon-condensate-dynamics-FRW}, an example of a
gluon-condensate-induced modification of gravity is presented
and the corresponding field equations are derived, which are then reduced
for the case of a spatially flat Friedmann--Robertson--Walker universe.
In Sec.~\ref{sec:Three-component-universe}, the resulting evolution of
a simple three-component model universe is studied both analytically
and numerically, in order to establish whether or not a model universe
can be obtained which resembles the observed ``accelerating Universe''
~\cite{Riess-etal1998,Perlmutter-etal1998}.
In Sec.~\ref{sec:Conclusion},  concluding remarks are presented.

\section{QCD--scale modified gravity and cosmology}
\label{sec:Gluon-condensate-dynamics-FRW}

\subsection{Theory: Action and field equations}
\label{subsec:Theory}

It has been argued~\cite{KlinkhamerVolovik2009a} that, in a de-Sitter
universe with Hubble constant $H$, a QCD--scale vacuum energy density
$\rho_{V} \sim |H|\,\Lambda_\text{QCD}^3$ could arise
from infrared effects of the gluon propagator.
Since the de-Sitter universe has Ricci curvature scalar $|R| \sim H^2$
and the particular gluon condensate $q$ has energy scale
$q \sim \Lambda_\text{QCD}^4$,
one is led to consider the following modified-gravity action
($\hbar=c=1$):\bsubeqs\label{eq:action-S-f}\beqa
S_\text{eff}&=&
\int_{\mathbb{R}^4} \,d^4x\, \sqrt{-g}\;
\Big[ K\, \widetilde{f}(R,q) +\epsilon(q)+\mathcal{L}_{M}(\psi)\Big]\,,
\label{eq:action-S}\\[2mm]
\widetilde{f}&\equiv& R + \widetilde{h}
              \equiv  R + \eta\,K^{-1}\,|R|^{1/2}\,|q|^{3/4} \,,
\label{eq:action-f}
\eeqa
\esubeqs
with gravitational coupling constant $K\equiv(16\pi G)^{-1}>0$,
dimensionless coupling constant $\eta > 0$
[standard general relativity has $\eta= 0\,$],
energy density $\epsilon(q)$ of the gluon condensate $q(x)$, and
matter field $\psi(x)$ [later on, this single matter component
will be generalized to $N$ matter components].
The precise definition of the gluon-condensate variable $q(x)$
in the context of QCD has been
given in Ref.~\cite{KlinkhamerVolovik2009a}, to which the reader is referred
for details. In the following, $q$ is simply assumed to be nonzero
and is, in fact, taken to be positive. The relation between the
gravitational constant $G$ and Newton's constant
$G_{N}$~\cite{Cavendish1798,MohrTaylorNewell2008}
will be discussed in Sec.~\ref{subsec:Analytic-results}.
Throughout, the conventions of Ref.~\cite{Weinberg1972} are used, in
particular, those for the Riemann tensor and the metric signature $(-+++)$.

The field equations from \eqref{eq:action-S-f} are fourth order
and it is worthwhile to switch to the scalar-tensor formulation which has
field equations of second order.
The equivalent Jordan-frame Brans--Dicke
theory~\cite{Weinberg1972,BransDicke1961,Will1993,Uzan2003} has action
\bsubeqs\label{eq:BDaction-S-U}
\beqa
S_\text{eff}^\text{\,(BD)}&=&
\int_{\mathbb{R}^4} \,d^4x\, \sqrt{-g}\;
\Big[ K\, \Big( \phi\, R - U(\phi,q) \Big)
+\epsilon(q)+\mathcal{L}_{M}(\psi)\Big]\,,
\label{eq:BDaction-S}\\[2mm]
U&\equiv& -(1/4)\, (\eta^2/K^2)\,|q|^{3/2}/(1-\phi) \,,
\label{eq:BDaction-U}
\eeqa
\esubeqs
in terms of a dimensionless scalar field $\phi$ restricted to
values less than $1$ [$\phi$ would be greater than $1$
for the $\eta < 0$ case not considered here].
The $\phi$ dependence of potential \eqref{eq:BDaction-U} allows for
the so-called chameleon effect~\cite{KhouryWeltman2004},
which will be briefly discussed at the end of this subsection.\footnote{See
also Ref.~\cite{MotaBarrow2004} for chameleon-type effects in a different context
and Ref.~\cite{Tamaki_etal2008} for recent analytic and
numerical work on the scalar profiles from compact objects,
extending the original analysis of Ref.~\cite{KhouryWeltman2004}.}
The proof of the classical equivalence of the actions
\eqref{eq:action-S-f} and \eqref{eq:BDaction-S-U}, for $\eta \ne 0$ and $q \ne 0$,
is not affected by the presence of the $q$--field in the function $\widetilde{f}$
of \eqref{eq:action-f}.
See, e.g., Refs.~\cite{Faulkner-etal2007,Brax-etal2008,SotiriouFaraoni2008}
for details of the proof, which is straightforward and need not be
repeated here. Anyway, the classical equivalence of \eqref{eq:action-S-f}
and \eqref{eq:BDaction-S-U} can be verified directly
by eliminating $\phi$ from \eqref{eq:BDaction-S}, using its
field equation $R=\partial U/\partial\phi$ with $U(\phi)$
given by \eqref{eq:BDaction-U}.

At this moment, two remarks may be helpful to place the theory considered
in context.
First, the rigorous microscopic derivation of the
effective action \eqref{eq:action-S-f} remains a major outstanding problem,
because only a rough argument has been given in the appendix
of Ref.~\cite{KlinkhamerVolovik2009a}, where $\eta$ was called $f$
(see also Ref.~\cite{ThomasUrbanZhitnitsky2009} for a general discussion
of the physics involved and \cite{endnote-heuristics} for
a heuristic argument).  Awaiting this derivation,
the main motivation of \eqref{eq:action-S-f} is that it naturally gives
the correct order of magnitude for the present vacuum energy density
(see Ref.~\cite{KlinkhamerVolovik2009a} and Sec.~\ref{sec:Conclusion}).
Just to be crystal clear: the term $\widetilde{h}$ in \eqref{eq:action-f}
is, at present, purely hypothetical
and the aim of this article is to explore its cosmological consequences,
leaving aside its theoretical derivation.

Second, the effective action \eqref{eq:action-S-f}
is only considered to be valid on cosmological length scales and
additional nonstandard terms in $\widetilde{f}(R,q)$ can be expected to be
operative at smaller length scales, relevant to solar-system
tests and laboratory experiments~\cite{Faulkner-etal2007,Brax-etal2008}.
Purely phenomenologically, the $\widetilde{h}$
term in \eqref{eq:action-f} could, for example, be replaced by an extended term
\beq\label{eq:h_ext}
\widetilde{h}_\text{ext}=
\eta\,\,K^{-1}\,|q|^{9/4}\,|R|^{1/2} \big/\big(|q|^{3/2}+\zeta\, K^2|R|\big)\,,
\eeq
with constants $0<\eta\ll |\zeta| \lesssim 1$.
This term $\widetilde{h}_\text{ext}$
vanishes as $|R|^{-1/2}$ at large enough curvatures and,
for $\eta \sim  10^{-3}$  and $|\zeta| \sim 1$,
is consistent with the relevant bound in Ref.~\cite{Brax-etal2008}
based on the E\"{o}t--Wash laboratory experiment~\cite{Kapner-etal2006}.

Returning to the action \eqref{eq:BDaction-S-U},
the field equations are obtained from the variational principle for
variations $\delta g_{\mu\nu}$ of the metric $g_{\mu\nu}$,
variations  $\delta\phi$ of the Brans--Dicke field $\phi$,
and variations  $\delta A$ of the
microscopic field $A$ responsible for $q$ condensate (see, in
particular, Refs.~\cite{KlinkhamerVolovik2008b,KlinkhamerVolovik2009a}).
Specifically, the field equations are
\bsubeqs\label{eq:BDfield-eqs-Gmunu-R-mu-tmp}
\beqa
R^{\mu\nu}-\frac{1}{2}\,R\,g^{\mu\nu}
&=&
-\frac{1}{2\phi\,K}\,
\Big( T_{M}^{\mu\nu} -\widetilde{\epsilon}\, g^{\mu\nu}\Big)
-\frac{1}{2\phi}\,\widetilde{U}\,g^{\mu\nu}
-\frac{1}{\phi}\,\Big( \nabla^\mu\nabla^\nu- g^{\mu\nu}\,\Box\Big)\phi\,,
\label{eq:BDfield-eqs-Gmunu-tmp}\\[2mm]
R&=&\frac{\partial U}{\partial\phi} \,,
\label{eq:BDfield-eqs-R-tmp}
\\[2mm]
\frac{\partial\epsilon}{\partial q}-K\,\frac{\partial U}{\partial q} &=& \mu \,,
\label{eq:BDfield-eqs-mu-tmp}
\eeqa
\esubeqs
with the covariant derivative $\nabla_\mu$,
the invariant d'Alembertian $\Box \equiv \nabla^\nu \nabla_\nu$,
the energy-momentum tensor $T_{M}^{\mu\nu}$ of the matter field $\psi$,
the integration constant $\mu$, and the effective energy densities
\bsubeqs\label{eq:widetilde-epsilon-U}
\beqa
\widetilde{\epsilon} &\equiv& \epsilon -q\,\frac{\partial\epsilon}{\partial q}\,,
\label{eq:widetilde-epsilon}\\[1mm]
\widetilde{U} &\equiv& U -q\,\frac{\partial U}{\partial q}\,.
\label{eq:widetilde-U}
\eeqa
\esubeqs
Two comments are in order.
First, the reason of having the extra term
$-q\,\partial\epsilon/\partial q$ in \eqref{eq:widetilde-epsilon}
and $-q\,\partial U/\partial q$ in \eqref{eq:widetilde-U} is the fact that the
field $q$ is not fundamental but contains, in addition to the
microscopic field $A$ mentioned above, the inverse metric $g^{\mu\nu}$
(see Sec. II of Ref.~\cite{KlinkhamerVolovik2009a}).
Second, the constant $\mu$ on the right-hand side of \eqref{eq:BDfield-eqs-mu-tmp}
can be interpreted, for spacetime-independent $q$ and $dU/dq=0$,
as the chemical potential corresponding to the conserved charge $q$
(see, in particular, the detailed discussion in Secs.~II A and B
of Ref.~\cite{KlinkhamerVolovik2008a}).

For completeness, also the generalized Klein--Gordon equation is given,
which is obtained by taking the trace of \eqref{eq:BDfield-eqs-Gmunu-tmp} and
using \eqref{eq:BDfield-eqs-R-tmp}:
\beq
\Box\, \phi =
\frac{1}{6\,K}\,\Big( T_{M} -4\,\widetilde{\epsilon}\Big)
+\frac{2}{3}\,\widetilde{U} -\frac{1}{3}\,\phi\,\frac{\partial U}{\partial \phi}\,,
\label{eq:BDfield-eqs-Box-eta-tmp}
\eeq
with the matter energy-momentum trace
$T_{M}\equiv T_{M}^{\mu\nu}\,g_{\mu\nu}$.

Eliminating $q\,\partial U/\partial q$ from \eqref{eq:BDfield-eqs-Gmunu-tmp}
and \eqref{eq:BDfield-eqs-mu-tmp}, the final field equations are
\bsubeqs\label{eq:BDfield-eqs-Gmunu-R-drhoVdq}
\beqa
R^{\mu\nu}-\frac{1}{2}\,R\,g^{\mu\nu}
&=&
-\frac{1}{2\phi\,K}\,
\Big( T_{M}^{\mu\nu} -\rho_{V}\, g^{\mu\nu}\Big)
-\frac{1}{2\phi}\,U\,g^{\mu\nu}
-\frac{1}{\phi}\,\Big( \nabla^\mu\nabla^\nu- g^{\mu\nu}\,\Box\Big)\phi\,,
\label{eq:BDfield-eqs-Gmunu}\\[2mm]
R&=&\frac{\partial U}{\partial\phi} \,,
\label{eq:BDfield-eqs-R}
\\[2mm]
\frac{\partial\rho_{V}}{\partial q} &=& K\,\frac{\partial U}{\partial q}\,,
\label{eq:BDfield-eqs-drhoVdq}
\eeqa
\esubeqs
in terms of the gravitating vacuum energy density
\beq
\rho_{V}(q)\equiv \epsilon(q) -\mu\, q\,,
\label{eq:EinsteinFRW-rhoV}
\eeq
with the integration constant $\mu$. Equally, the generalized Klein--Gordon
equation \eqref{eq:BDfield-eqs-Box-eta-tmp} becomes
\beq
\Box\, \phi =
\frac{1}{6\,K}\,\Big( T_{M} -4\,\rho_{V}\Big)
+\frac{2}{3}\,U -\frac{1}{3}\,\phi\,\frac{\partial U}{\partial \phi}\,,
\label{eq:BDfield-eqs-Box-eta}
\eeq
where the very last term on the right-hand side, in particular,
is relevant to the previously mentioned chameleon effect.
With \eqref{eq:BDfield-eqs-R}, this last term of \eqref{eq:BDfield-eqs-Box-eta}
becomes $(-R/3)\,\phi$ and corresponds to an effective mass square term
for the scalar field, with a mass square of the order of
$\rho_{M}/K$ for the case of a pressureless perfect fluid.
This is indeed one aspect of the chameleon effect, namely, an effective
mass value dependent on the environment~\cite{KhouryWeltman2004}.

\subsection{Differential equations for a flat FRW universe}
\label{subsec:FRW-equations}

For a spatially flat ($k=0$) Friedmann--Robertson--Walker (FRW)
universe~\cite{Weinberg1972} with scale factor $a(\dimfultime)$
and matter described by a perfect fluid,
the $00$ and $11$ components of the  generalized Einstein field equation
\eqref{eq:BDfield-eqs-Gmunu} can be combined to give a
generalized Friedmann equation.
Together with equations obtained directly from \eqref{eq:BDfield-eqs-R}
and \eqref{eq:BDfield-eqs-Box-eta}, the relevant equations are then
\bsubeqs\label{eq:field-eqs-FRW-phidot-Hdot-phiddot}
\beqa
H^2\,\phi
&=&
\frac{1}{6\,K}\,\rho_\text{tot}-\frac{1}{6}\,U- H\, \dot{\phi}\,,
\label{eq:field-eqs-FRW-phidot-second}\\[2mm]
\dot{H} &=&-2H^2- \frac{1}{6}\, \frac{\partial U}{\partial \phi}\,,
\label{eq:field-eqs-FRW-Hdot}\\[2mm]
\ddot{\phi}  &=&
-3H\,\dot{\phi}
+\frac{1}{6\,K}\,\Big( \rho_\text{tot}-3\, P_\text{tot}\Big)
-\frac{2}{3}\,U +\frac{1}{3}\,\phi\,\frac{\partial U}{\partial \phi}\,,
\label{eq:field-eqs-FRW-phiddot}
\eeqa
\esubeqs
with the overdot standing for the derivative with respect to $\dimfultime$
(the somewhat unusual notation $\dimfultime$ is used
for the dimensionful cosmic time, in order to reserve the letter $\dimlesstime$
for the dimensionless time later on).
The total energy density and pressure are given by
\bsubeqs \beq\label{eq:rho-total}
\rho_\text{tot}\equiv \rho_{V}+\rho_{M}\,,\quad P_\text{tot} \equiv
P_{V}+P_{M}\,,
\eeq
for the gravitating vacuum energy density
\beq
\rho_{V}(q)= -P_{V}(q) = \epsilon(q) -\mu\, q\,,
\label{eq:EinsteinFRW-rhoV-PV}
\eeq
\esubeqs
as discussed in the previous subsection.
Observe that \eqref{eq:field-eqs-FRW-phidot-second} reproduces
the standard Friedmann equation for $U=0$, $\phi=1$, and
$K\equiv(16\pi G)^{-1}=(16\pi G_{N})^{-1} \equiv  K_{N}$.

The last two equations in \eqref{eq:field-eqs-FRW-phidot-Hdot-phiddot}
are, respectively, first- and second-order ordinary differential equations
(ODEs) for $H$ and $\phi$. Two further ODEs can be obtained as follows.
First, multiplying \eqref{eq:BDfield-eqs-drhoVdq}
by $\dot{q}$ gives an equation for the time dependence of the
vacuum energy density,
\bsubeqs
\beq
\dot{\rho}_{V} =
K\,\left(\dot{U}-\dot{\phi}\;\frac{\partial U}{\partial \phi}\right)\,,
\label{eq:field-eqs-FRW-rhoVdot}
\eeq
which describes the energy exchange between the vacuum and the
nonstandard gravitational field ($U\ne 0$).
Second, the standard energy conservation of matter gives
\beq
\dot{\rho}_{M} =  -3H\,\Big(\rho_{M}+P_{M}\Big)
                    =   -3H\,\Big(1+w_{M}\Big)\,\rho_{M}\,,
\label{eq:matter-energy-conservation}
\eeq
\esubeqs
where the matter equation-of-state (EOS)
parameter $w_{M}\equiv P_{M}/\rho_{M}$ has been introduced
(henceforth, $w_{M}$ will be
assumed to be time independent). Equation \eqref{eq:matter-energy-conservation}
implies that, for the theory considered,
there is no energy exchange between vacuum and matter
(such an energy exchange for a different version of $q$--theory
has been studied in Ref.~\cite{Klinkhamer2008}).

\subsection{Dimensionless variables and ODEs}
\label{subsec:Dimensionless-equations}

Now rewrite the cosmological equations in appropriate  microscopic  units.
The gluon condensate $q$ from
Refs.~\cite{ShifmanVainshteinZakharov1978,KlinkhamerVolovik2009a} has
the dimension of energy density, $[q]=[\epsilon]$, which implies that the
corresponding  integration constant $\mu$ is dimensionless, $[\mu]=[1]$.
The equilibrium value $q_{0}$ of the gluon-condensate variable $q$
is taken to be determined by a laboratory experiment in an environment
with negligible spacetime curvature and has the order of magnitude
$q_{0}\equiv E_\text{QCD}^4 =\text{O}(10^{9}\,\text{eV}^4)$;
see Sec.~\ref{subsec:Exploratory-numerical-results} for further remarks.
 From this moment on, consider $N$ matter components, labeled by an
index $n=1, \ldots , N$.

Specifically, the following dimensionless variables
$\dimlesstime$, $h$, $f$, $r$, $u$, and $\dimlessscalar$ can be introduced:
\bsubeqs\label{eq:Dimensionless-var}
\begin{align}
\hspace*{-.8cm}
\dimfultime&\equiv \dimlesstime \;K\big/q_{0}^{3/4}\,,
&H(\dimfultime)&\equiv h(\dimlesstime)\;q_{0}^{3/4}\big/K\,,
\label{eq:Dimensionless1-tau-h}
\\[2mm]
\hspace*{-.8cm}
q(\dimfultime)&\equiv f(\dimlesstime)\; q_{0}\,,
&\rho(\dimfultime) &\equiv r(\dimlesstime)\;q_{0}^{3/2}\big/K\,,
\label{eq:Dimensionless1-f-rS}
\\[2mm]
\hspace*{-.8cm}
U(\dimfultime)&\equiv u(\dimlesstime)\;q_{0}^{3/2}\big/K^2\,,
&\phi(\dimfultime)&\equiv \dimlessscalar(\dimlesstime)\,.
\label{eq:Dimensionless1-u-s}
\end{align}
\esubeqs
Observe that all dimensionless quantities are denoted by lower-case Latin letters.
A further rescaling $t=t^\prime/\eta$ and $h=h^\prime\,\eta$
will not be used in the present article, as the effects from the
unknown coupling constant $\eta$ are preferred to be kept as explicit as
possible.

It is, then, straightforward to obtain the dimensionless versions of the
algebraic equation \eqref{eq:BDfield-eqs-drhoVdq}, the last two ODEs
in \eqref{eq:field-eqs-FRW-phidot-Hdot-phiddot},
and the matter conservation equation \eqref{eq:matter-energy-conservation}
generalized to $N$ matter components. This gives
a closed system of $4+N$ equations for the $4+N$
dimensionless variables $f(\dimlesstime)$, $h(\dimlesstime)$,
$\dimlessscalar(\dimlesstime)$, $v(\dimlesstime)$, and $r_{M,n}(\dimlesstime)$.
Specifically, this system of equations consists of a single algebraic equation,
\beqa
\hspace*{-5mm}
\frac{\partial r_{V}(f)}{\partial f} &=&
\frac{\partial u(s,f)}{\partial f}\,,
\label{eq:alg-eq-FRWdim-f}
\eeqa
and $3+N$ ODEs,
\bsubeqs\label{eq:4ODEsFRWdim}
\beqa
\hspace*{-5mm}
\dot{h} &=& -2\,h^2 - \frac{1}{6}\,\frac{\partial u}{\partial\dimlessscalar}\,,
\label{eq:4ODEsFRWdim-h}\\[2mm]
\hspace*{-5mm}
\dot{s} &=& v\,,
\label{eq:4ODEsFRWdim-s}\\[2mm]
\hspace*{-5mm}
\dot{v} &=&
\frac{1}{6}\,\big(r_\text{tot}-3\, p_\text{tot}\big)-3\,h\,v-  \frac{2}{3}\,u
+ \frac{1}{3}\,\dimlessscalar\,\frac{\partial u}{\partial \dimlessscalar}\,,
\label{eq:4ODEsFRWdim-v}\\[2mm]
\hspace*{-5mm}
\dot{r}_{M,n}
&=&
-3\,h \,\big(1+w_{M,n}\big)\,\,r_{M,n}\,,
\label{eq:4ODEsFRWdim-rM}
\eeqa \esubeqs
where, now, the overdot stands for differentiation with respect to
the dimensionless cosmic time $\dimlesstime$
and the dimensionless total energy density and pressure are given by
\bsubeqs\label{eq:rtot-ptot}
\beqa
\hspace*{-0mm}
r_\text{tot} &=& +r_{V}+ \sum_{n=1}^{N}\; r_{M,n}\,,
\label{eq:rtot}\\[2mm]
\hspace*{-0mm}
p_\text{tot} &=& -r_{V}+\sum_{n=1}^{N}\; w_{M,n}\,r_{M,n}\,,
\label{eq:ptot}
\eeqa
\esubeqs
with matter EOS parameters $w_{M,n}$ still to be specified.
The dimensionless vacuum energy density $r_{V}$ appearing in the above
equations will be discussed in Sec.~\ref{subsec:Ansatz-rV-solution-f(s)}.
The dimensionless potential $u$ has already been defined by
\eqref{eq:BDaction-U} and \eqref{eq:Dimensionless1-u-s}, but
will be given again in Sec.~\ref{subsec:Ansatz-rV-solution-f(s)}.

With the solution of Eqs.~\eqref{eq:alg-eq-FRWdim-f}--\eqref{eq:4ODEsFRWdim}
for appropriate boundary conditions,
it is possible to verify \emph{a posteriori} the Friedmann-type equation
\eqref{eq:field-eqs-FRW-phidot-second} in dimensionless form:
\beq
 h^2\,\dimlessscalar+h\,v = \big(r_\text{tot} - u\big)\big/ 6  \,,
\label{eq:Friedmann-type-eq}
\eeq
which, in general, is guaranteed to hold by the contracted Bianchi identities and
energy conservation (cf. Refs.~\cite{Weinberg1972,Klinkhamer2008}).
Specifically, if the solution of
Eqs.~\eqref{eq:alg-eq-FRWdim-f}--\eqref{eq:4ODEsFRWdim}
satisfies \eqref{eq:Friedmann-type-eq} at one particular time,
then \eqref{eq:Friedmann-type-eq} is satisfied at all the times considered.
The additional constraint \eqref{eq:Friedmann-type-eq} will
provide a valuable check on the numerical solution of the equations.

\subsection{Ansatz for $\boldsymbol{r_{V}(f)}$ and solution for $\boldsymbol{f(s)}$}
\label{subsec:Ansatz-rV-solution-f(s)}

The only further input needed for the cosmological
Eqs.~\eqref{eq:alg-eq-FRWdim-f}--\eqref{eq:4ODEsFRWdim}
is an \emph{Ansatz} for the gravitating vacuum energy density
$\rho_{V}(q)$ from \eqref{eq:EinsteinFRW-rhoV} or the corresponding
dimensionless quantity $r_{V}$ from \eqref{eq:Dimensionless1-f-rS}.
In Refs.~\cite{KlinkhamerVolovik2008a,KlinkhamerVolovik2008b,KlinkhamerVolovik2008c,KlinkhamerVolovik2009a},
it was argued that the vacuum variable $q$ of the late Universe is close to its
flat-spacetime equilibrium value $q_{0}$ and the quadratic approximation can be used
\beq
r_{V} =\gamma\, (1-f)^2\,,
\label{eq:rV-Ansatz}
\eeq
with positive constant $\gamma$.

 From the $r_{V}$ definition in \eqref{eq:Dimensionless1-f-rS},
the constant $\gamma$ in \eqref{eq:rV-Ansatz} can be expected to be of
order $Z^{-1}$, with definition
\beq
Z \equiv  q_{0}^{1/2}\;K^{-1}
  \sim    16\pi\;\big(E_\text{QCD}/E_\text{Planck}\big)^2
  \sim   10^{-38}\,,
\label{eq:Z-definition}
\eeq
for the quantum-chromodynamics energy scale $E_\text{QCD} \approx 0.2 \;\text{GeV}$
and the standard gravitational energy scale
$E_\text{Planck} \equiv \sqrt{\hbar\, c^5/G_{N}}
\approx 1.22 \times 10^{19}\;\text{GeV}$ (having set $G \sim G_{N}$;
see Sec.~\ref{subsec:Analytic-results}). According to the discussion in
Refs.~\cite{KlinkhamerVolovik2008a,KlinkhamerVolovik2008b,KlinkhamerVolovik2008c,KlinkhamerVolovik2009a},
$f$ can also be expected to be sufficiently close to $1$, in order to reproduce
an $r_{V}$ value of order unity or less for the present Universe.
For technical reasons, the value $Z=10^{-2}$ is taken in a first numerical
study (Sec.~\ref{subsec:Exploratory-numerical-results}).
Later, the proper boundary conditions and scaling behavior are considered
(Sec.~\ref{subsec:Elementary-scaling-analysis}).

The dimensionless scalar potential $u(\dimlessscalar,f)$ from \eqref{eq:BDaction-U}
and \eqref{eq:Dimensionless1-u-s} can be written as
\beq
u(t) \equiv U\,K^2\,q_{0}^{-3/2}
      =     -(\eta^2/4)\;\frac{f(t)^{3/2}}{1-\dimlessscalar(t)}\,,
\label{eq:dimensionless-potential-u}
\eeq
where a relatively small value for $\eta$ appears to be
indicated~\cite{KlinkhamerVolovik2009a} by the
measured value of the vacuum energy density;
see Secs.~\ref{subsec:Analytic-results} and \ref{subsec:Elementary-scaling-analysis}
for further discussion on the numerical value of $\eta$.

With the specific functions
\eqref{eq:rV-Ansatz} and \eqref{eq:dimensionless-potential-u},
Eq.~\eqref{eq:alg-eq-FRWdim-f} is a quadratic in $\sqrt{f}$ and
the positive root gives
\bsubeqs\label{eq:fsolution-B-zeta}
\beqa
\overline{f}_{\pm}(s)     &=&   \left( \sqrt{1+D(s)^2} \pm D(s) \,\right)^2\,,
\label{eq:fsolution}\\[2mm]
D(s)     &\equiv& \kappa/|1-s| \geq 0\,,
\label{eq:C}\\[2mm]
\kappa &\equiv& (3/32)\, \eta^2/\gamma \geq 0\,,
\label{eq:kappa}
\eeqa
\esubeqs
where the minus sign inside the outer parentheses on the right-hand side
of \eqref{eq:fsolution} holds for $s<1$
[the plus sign appears for the $s>1$ case not considered here].
Expression \eqref{eq:fsolution}
can then be used to eliminate all occurrences of $f$
in the $3+N$ ODEs~\eqref{eq:4ODEsFRWdim} for the remaining $3+N$ variables
$h(\dimlesstime)$, $\dimlessscalar(\dimlesstime)$, $v(\dimlesstime)$,
and $r_{M,n}(\dimlesstime)$.
Referring to the ODEs~\eqref{eq:4ODEsFRWdim} in the following, it
will be understood that $f$ has been replaced by $\overline{f}_{-}(s)$
from \eqref{eq:fsolution-B-zeta}.

\section{Three-component model universe}
\label{sec:Three-component-universe} \vspace*{0mm}

\subsection{Preliminaries}
\label{subsec:Preliminaries}

The modified-gravity theory considered in this article
has been presented in Sec.~\ref{subsec:Theory}
and the corresponding dynamical equations for a spatially flat FRW universe
in Secs.~\ref{subsec:FRW-equations}--\ref{subsec:Ansatz-rV-solution-f(s)}.
The specific model studied in this section is a simplified version with
only three components labeled $n=0,1,2$:
\begin{enumerate}
\setcounter{enumi}{-1}
\item
A gluon condensate [described by the dimensionless variable $f$]
with dimensionless energy density $r_{V}(f)$ from \eqref{eq:rV-Ansatz} and
constant  equation-of-state parameter $w_{V}=-1$, which
is taken to give rise to a nonanalytic term in the modified-gravity
action \eqref{eq:action-S-f}.
\item
A perfect fluid of ultrarelativistic matter [e.g., photons]
with energy density $r_{M,1}$ and constant EOS parameter $w_{M,1}=1/3$.
\item
A perfect fluid of nonrelativistic matter
[e.g., cold dark matter (CDM) and baryons (B)]
with energy density $r_{M,2}$ and constant EOS parameter $w_{M,2}=0$.
\end{enumerate}
From the scalar-tensor formalism of the gluon-condensate-induced modification
of gravity, there is also the auxiliary Brans--Dicke scalar $\dimlessscalar(t)$
to consider, with the dimensionless potential $u(\dimlessscalar,f)$
from \eqref{eq:dimensionless-potential-u}.

The relevant ODEs follow from \eqref{eq:4ODEsFRWdim}
by letting the matter label run over $n=1,2$.
The ideal starting point of the calculations would be some time after
the QCD crossover at $T \sim \Lambda_\text{QCD}$ with
$r_{M,1} \gg r_{M,2}$.
The physical idea is that the expansion of the Universe was standard up till that
time and that, then, a type of phase transition occurred with the creation
of the gluon condensate. Clearly, the gluon condensate can be expected to
start out in a nonequilibrium state, $f \ne 1$ and $s \ne 1$.
These issues will be discussed further in
Sec.~\ref{subsec:Elementary-scaling-analysis}.

At this moment, it is useful to recall the basic equations of a standard
flat FRW universe~\cite{Weinberg1972,Weinberg2008}
with gravitational coupling constant $G=G_{N}$ or $K=K_{N}$.
For two components, a pressureless material fluid labeled $M$
and an unknown fluid labeled $X$, these equations are
\bsubeqs\label{eq:standard-FRW-dota-ddota}
\beqa
6\,h^2 &\equiv& 6\,(\dot{a}/a)^2 = r_{M} + r_{X}\,,
\label{eq:standard-FRW-dota}\\[2mm]
-12\,\ddot{a}/a &=& r_{M} + r_{X} + 3\,p_{M} + 3\,p_{X}
                          = 
r_{M} + r_{X}\,\big(1 + 3\,w_{X}\big)\,,
\label{eq:standard-FRW-ddota}
\eeqa \esubeqs
where $p_{M}$ in \eqref{eq:standard-FRW-ddota} has been
set to zero and the EOS parameter
$w_{X}\equiv p_{X}/r_{X}$ has been introduced.
The standard energy-density parameters are defined as follows:
\bsubeqs\label{eq:standard-OmegaMX-wX}
\beqa
\Omega_{M} &\equiv& r_{M}/(6\,h^2)\,,\quad
\Omega_{X} \equiv r_{X}/(6\,h^2) = 1-\Omega_{M}\,.
\label{eq:standard-OmegaMX}
\eeqa
In addition, the following combination of observables can be
introduced to determine the unknown EOS parameter:
\beq
\overline{w}_{X}
\equiv
-\frac{2}{3}\,\left(\frac{\ddot{a}\,a}{(\dot{a})^2}+\frac{1}{2}\right)\;
 \frac{1}{1-\Omega_{M}}
=
w_{X}\,,
\label{eq:standard-wX}
\eeq \esubeqs
where the last equality holds, again, for $p_{M}=0$.
See, e.g., Refs.~\cite{WellerAlbrecht2002,SahniStarobinsky2006}
for details on how to reconstruct the dark-energy equation of state
from observations.

In order to be specific, take the following fiducial values:
\beq
\big\{ \Omega_{M},\, \Omega_{X},\, \overline{w}_{X}
\big\}^{\text{standard\;FRW}}_{\text{present}}
=
\big\{ 0.25,\,0.75,\, -1 \big\}\,,
\label{eq:FRW-OmegaXM-wX}
\eeq
which agree more or less with the recent data compiled in
Refs.~\cite{Freedman2001,Eisenstein2005,Astier2006,Riess2007,Komatsu2008,Vikhlinin-etal2008}.
The standard flat FRW universe with parameters \eqref{eq:FRW-OmegaXM-wX}
corresponds, in fact, to the basic $\Lambda$CDM model~\cite{Weinberg2008} with
CDM energy density $r_{M}\propto 1/a^3$
(with constant EOS parameter $w_{M}=0$) and
time-independent vacuum energy density $l \equiv r_{X}\propto a^0$
(with constant EOS parameter $w_{X}=-1$ and $l$ the dimensionless
version of the cosmological constant $\Lambda$).

Returning to the modified-gravity theory
\eqref{eq:action-S-f}--\eqref{eq:BDaction-S-U}, the same observables
$\Omega$ and $\overline{w}_{X}$ can be identified.
Specifically, the generalized
Friedmann equation \eqref{eq:Friedmann-type-eq}  gives
\bsubeqs\label{eq:Omegabar}
\beqa
\Omega_{X}+\Omega_{M}&=&1\,,
\label{eq:Omegabar-XplusM}\\[2mm]
\Omega_{X}   &\equiv& \Omega_\text{grav}+\Omega_{V}\,,
\label{eq:Omegabar-X}\\[2mm]
\Omega_\text{grav}&\equiv& 1-s-\dot{s}/h-u/(6h^2)\,,
\label{eq:Omegabar-grav}\\[2mm]
\Omega_{V} &\equiv& r_{V}/(6h^2)\,,
\label{eq:Omegabar-V}\\[2mm]
\Omega_{M} &\equiv& r_{M}/(6h^2)\,,
\label{eq:Omegabar-M}
\eeqa
\esubeqs
where $\Omega_\text{grav}$ is the new ingredient, as it
vanishes for the standard theory with $u=0$ and $s=1$.
Similarly, the effective EOS parameter
of the unknown component $X$ can be extracted
from \eqref{eq:4ODEsFRWdim} and \eqref{eq:Friedmann-type-eq} for $p_{M}=0$:
\beqa\label{eq:modgrav-wXbar}
\overline{w}_{X}
&\equiv&
-\frac{2}{3}\,\left(\frac{\ddot{a}\,a}{(\dot{a})^2}+\frac{1}{2}\right)\;
 \frac{1}{1-\Omega_{M}}
           = 
 -\;\frac{r_{V} - u -4\,h\,\dot{s} -2\,\ddot{s} \phantom{\;(1-s)}}
         {r_{V} - u -6\,h\,\dot{s} +r_{M}\,(1-s) )}\;.
\eeqa
The right-hand side of \eqref{eq:modgrav-wXbar} shows that
$\overline{w}_{X}$ of the modified-gravity model \eqref{eq:BDaction-S-U}
approaches the value $-1$ in the limit of vanishing
matter content and constant Brans--Dicke scalar $s$ as $t\to\infty$.
\emph{A priori\,}, there is no reason why this approach cannot be from
below, so that $1+\overline{w}_{X}$ would be negative for a while
(cf. Ref.~\cite{CarrollDeFeliceTrodden2004}).

The main goal of this section is to get a quasirealistic model
for the ``present universe,''
which is taken to be defined by a value of approximately $0.25$
for the matter energy-density parameter $\Omega_{M}$.
This can only be done with a numerical solution of the
ODEs, but, first, analytic results relevant to the
asymptotic behavior at early and late times are discussed.

\subsection{Analytic results}
\label{subsec:Analytic-results}

It is not difficult to get two types of analytic solutions of the
combined ODEs~\eqref{eq:4ODEsFRWdim} and \eqref{eq:Friedmann-type-eq}
for the specific functions \eqref{eq:rV-Ansatz}
and \eqref{eq:dimensionless-potential-u}, having used
solution \eqref{eq:fsolution-B-zeta} to eliminate $f$ in favor of $s$.
The first corresponds to a Friedmann universe with relativistic matter
and without vacuum energy. The second corresponds to a de-Sitter-type universe
without  matter and with an effective form of vacuum energy.

For $\eta=0$, the first analytic solution of
\eqref{eq:4ODEsFRWdim}--\eqref{eq:fsolution-B-zeta}  has only
relativistic matter ($w_{M,1}=1/3$) contributing to the expansion.
Specifically, this Friedmann solution  (labeled ``$\text{F}$'')
is given by
\bsubeqs\label{eq:Fsolution}
\beqa
h^\text{(F)}             &=&  (1/2)\;\dimlesstime^{-1}\,,\quad
\dimlessscalar^\text{(F)}=f^\text{(F)}=1\,,
\\[2mm]
r_{M,1}^\text{(F)}  &=&  (3/2)\;\dimlesstime^{-2}\,,\quad
r_{M,2}^\text{(F)}   =  0\,.
\eeqa
\esubeqs
Remark that standard general relativity [which has, from the start, the action
equal to \eqref{eq:action-S-f} for $\eta=0$ and $G=G_{N}$]
allows for arbitrary values $r_{M,1}(1)$
and $r_{M,2}(1)$ at reference time $\dimlesstime=1$.

For $\eta > 0$, the second set of analytic solutions
of \eqref{eq:4ODEsFRWdim}--\eqref{eq:fsolution-B-zeta}
has only vacuum energy contributing to the expansion, together
with the effects of the gluon-condensate-induced modification of gravity
($\overline{w}_{X}=-1$).
This type of solution has constant (time-independent) variables
$h>0$ and $s \in (0,\,1)$, with $f$ given by \eqref{eq:fsolution}.
From \eqref{eq:4ODEsFRWdim-h} and \eqref{eq:4ODEsFRWdim-v},
using \eqref{eq:dimensionless-potential-u}, a cubic in $s$ is obtained,
which needs to be discussed first.

Specifically, the cubic in $x\equiv 1-s$ reads
\beq\label{eq:cubic}
9\,x^3 - 6\, x^2 + \big(1 + 9\,\kappa^2\big)\,x -6\, \kappa^2 =0\,,
\eeq
with parameter $\kappa$ defined by \eqref{eq:kappa}.
This cubic has three distinct real solutions
for $0<\kappa^2< (5\, \sqrt{5}-11)/18 \approx \big(0.100094\big)^2$.
Two of these solutions (with $2/3<s<1$) give stationary
de-Sitter-type solutions of the
ODEs~\eqref{eq:4ODEsFRWdim}--\eqref{eq:fsolution-B-zeta}.
These two roots can be written in manifestly real form by use of the
Chebyshev cube root
\bsubeqs\beqa
C_{1/3}(t)\,\Big|_{|t|<2} &\equiv& 2 \cos\big[(1/3) \arccos(t/2)\big]\,,\\
C_{1/3}(0)                &\equiv& \sqrt{3}\,.
\eeqa\esubeqs
Defining the auxiliary parameters
\bsubeqs\label{eq:parameters-p-q}
\beqa
p &\equiv& (1/3)\,\big(1/27+\kappa^2    \big)\,,\\[2mm]
q &\equiv& (2/9)\,\big(1/82-2\,\kappa^2 \big),,
\eeqa
\esubeqs
the relevant roots of \eqref{eq:cubic} are
\bsubeqs\label{eq:sroots}
\beqa
s_\text{high}   &=&
7/9 +\sqrt{p}\; \;C_{1/3}\big(\! -q\,p^{-3/2}\big)\,,
\\[2mm]
s_\text{mid} &=&
7/9 +\sqrt{p}\;\Big[ C_{1/3}\big(q\,p^{-3/2}\big)
-C_{1/3}\big(\!-q\,p^{-3/2}\big)\Big]\,,
\eeqa
\esubeqs
where the
third solution $s_\text{low}=7/3-s_\text{high}-s_\text{mid}$ can be omitted, as
it lies below $2/3$ for $\kappa$ in the domain considered
[the stationary limit of, e.g., Eq.~\eqref{eq:4ODEsFRWdim-v} requires $s\geq2/3$
because $r_{V}$ from \eqref{eq:rV-Ansatz} is non-negative by definition].

The first de-Sitter-type solution (labeled ``$\text{deS,0}$'' because
$f\sim 0$ for $|\kappa|\ll 1$) is then given by
\bsubeqs\label{eq:deSsolution0}
\beqa
s^\text{(deS,0)}
&=&
s_\text{high}
=
1 - 6\, \kappa^2 - 162\, \kappa^4 +\text{O}\big(\kappa^6\big)\,,
\\[2mm]
f^\text{(deS,0)}
&=&
\overline{f}_{-}\big( s_\text{high} \big)
=  9\, \big(\kappa^2 + 36\, \kappa^4\big) +\text{O}\big(\kappa^6\big)\,,
\\[2mm]
h^\text{(deS,0)}
&=&
\eta \big/\big(4\sqrt{3}\big)\;
\big|f^\text{(deS,0)}\big|^{3/4}\;\big|1-s^\text{(deS,0)}\big|^{-1}
           = 
\sqrt{\gamma/6}\;
\nonumber\\
&&\times
\left[ 1 - (81/2)\, \kappa^4 +\text{O}\big(\kappa^6\big)\right]\,,
\label{eq:deSsolution0-h}\\[2mm]
r_{M,n}^\text{(deS,0)}&=& 0\,,
\eeqa
\esubeqs
in terms of the function $\overline{f}_{-}(s)$ defined by \eqref{eq:fsolution}
and with an integer $n=1,2$ to label the different matter components.
Note that the expression in the middle of \eqref{eq:deSsolution0-h}
simply follows from \eqref{eq:4ODEsFRWdim-h} for $\dot{h}=0$ and $u$
from \eqref{eq:dimensionless-potential-u}.

The second solution (labeled ``$\text{deS,1}$'' because $f\sim 1$
for $|\kappa|\ll 1$) is given by
\bsubeqs\label{eq:deSsolution1}
\beqa
s^\text{(deS,1)}
&=&
s_\text{mid}
=
2/3 + \kappa + 3\, \kappa^2
+ (27/2)\, \kappa^3+ 81\, \kappa^4 +\text{O}\big(\kappa^5\big)\,,
\label{eq:deSsolution1-s}\\[2mm]
f^\text{(deS,1)}
&=&
\overline{f}_{-}\big( s_\text{mid} \big)
=1 - 6\, \kappa
- 27\, \kappa^3 - 162\, \kappa^4 +\text{O}\big(\kappa^5\big)\,,
\label{eq:deSsolution1-f}\\[2mm]
h^\text{(deS,1)}
&=&
\eta \big/\big(4\sqrt{3}\big)\;
\big|f^\text{(deS,1)}\big|^{3/4}\;\big|1-s^\text{(deS,1)}\big|^{-1}
= 
\sqrt{2\gamma\kappa}\big/ 1024 \;                      
\nonumber\\
&&\times \Big[ 1024 - 1536\, \kappa + 1152\, \kappa^2
+ 1728\, \kappa^3 + 17496\, \kappa^4
       +\text{O}\big(\kappa^5\big) \Big]\,,
\label{eq:deSsolution1-h}\\[2mm]
r_{M,n}^\text{(deS,1)}&=& 0\,,
\eeqa
\esubeqs
where $\kappa$ is non-negative according to the original definition \eqref{eq:kappa}.
Note that the last expressions of both \eqref{eq:deSsolution0-h}
and \eqref{eq:deSsolution1-h} are proportional
to $\sqrt{\gamma}$ with all further dependence
on $\gamma$ entering through the parameter $\kappa \propto \eta^2/\gamma$,
as can be expected on general grounds from
the ODEs~\eqref{eq:4ODEsFRWdim} without matter.

It is not quite trivial that there indeed exist de-Sitter-type
solutions in the modified-gravity theory \eqref{eq:action-S-f}.
The first solution \eqref{eq:deSsolution0} is far from the equilibrium
state $f_\text{equil}=1$
and the second solution \eqref{eq:deSsolution1} is close to it,
at least for $|\kappa|\ll 1$.
The scaling behavior of both solutions under the limit $\gamma\to\infty$
for constant $\eta$ is also different, with $h$ diverging for the
first solution and staying constant for the second.
For fixed parameters $\gamma$ and $\eta$,
numerical results suggest that the first solution
\eqref{eq:deSsolution0} is unstable and the second solution \eqref{eq:deSsolution1}
stable [and possibly an attractor].
In the following, the focus is on the second solution
close to the equilibrium value $f_\text{equil}=1$ (corresponding to $q=q_0$).

In fact, two remarks on the de-Sitter-type solution \eqref{eq:deSsolution1}
are in order. First, observe that local experiments in this model universe
with $\phi^\text{(deS,1)} \sim 2/3 < 1$
would have an increased effective gravitational coupling
\beqa\label{eq:Geff}
\overline{G}_{N} &\equiv&
G_\text{\,eff}^\text{\,local\;exps}\,\Big|^\text{(deS,1)}
\sim \big(1/\phi^\text{(deS,1)}\big)\; G \sim (3/2)\; G \,,
\eeqa
where the term $G/\phi^\text{(deS,1)}$ in the middle comes
directly from the combination $K\,\phi=\phi/(16\pi\, G)$ present in the
action \eqref{eq:BDaction-S-U}.
Here, ``local experiments'' denote experiments on length scales
very much less than the typical  length scale of de-Sitter-type spacetime,
the horizon distance $L_\text{hor}= c\,H^\text{(deS,1)}$, whose numerical
value will be discussed shortly.
It would then appear that the quantity \eqref{eq:Geff} must be
identified with Newton's gravitational constant $G_{N}$
as measured by Cavendish~\cite{Cavendish1798} and modern-day
experimentalists~\cite{MohrTaylorNewell2008}; see
\cite{endnote-G_Newton} for additional comments.

Second, the de-Sitter-type solution \eqref{eq:deSsolution1}
of model \eqref{eq:BDaction-S-U} or equivalently model \eqref{eq:action-S-f}
has the inverse Hubble constant
\beq\label{eq:hinverse}
\left(h^\text{(deS,1)}\right)^{-1}
=  4/\sqrt{3}\,\;\eta^{-1}
\approx
2.3 \times 10^{3}\;\left( \frac{10^{-3}}{\eta} \right)\,,
\eeq
as follows from \eqref{eq:deSsolution1-h} by neglecting terms suppressed by
powers of $\kappa=\text{O}(1/\gamma)=\text{O}(10^{-38})$ and anticipating
a particular order of magnitude for the model parameter $\eta$.
With the conversion factor from \eqref{eq:Dimensionless1-tau-h},
the dimensionless quantity \eqref{eq:hinverse} corresponds to
\beqa\label{eq:Hinverse}
\left(H^\text{(deS,1)}\right)^{-1}
&\sim&
4/\sqrt{3}\;\,\eta^{-1}\;(3/2)\,K_{N}\:q_{0}^{-3/4}
                        \sim 
8 \times 10^{17}\,\text{s}\,\left( \frac{10^{-3}}{\eta} \right) 
\left(\frac{200\;\text{MeV}}{q_{0}^{1/4}}\right)^3\;,
\eeqa
where, according to \eqref{eq:Geff}, an approximate  factor $3/2$ appears in going
from $K$ to the Newtonian value $K_{N}\equiv (16\pi G_{N})^{-1}$.
The time scale found in \eqref{eq:Hinverse}
is of the same order as the inverse Hubble constant
$(H_0)^{-1} 
            \approx  4.5\times 10^{17}\,\text{s}\;(0.70/h_{0})$
for the measured value $h_{0}\approx 0.70$ as reported in
Refs.~\cite{Freedman2001,Komatsu2008,Vikhlinin-etal2008}.

By equating the theoretical quantity
$1/H^\text{(deS,1)}$ from \eqref{eq:Hinverse}
multiplied by an \emph{ad hoc} factor $g=\half$
with the measured value $1/H_0$, a first estimate of the
model parameter $\eta$ in the original action \eqref{eq:action-S-f}
is obtained,
\beq\label{eq:eta-first-estimate}
\eta \sim \sqrt{3}\, K_{N}\:q_{0}^{-3/4} \:H_0 \sim 10^{-3}\,,
\eeq
for the $q_{0}$ and $H_0$ values mentioned in the previous paragraph.
Admittedly, the choice of one-half for the factor $g$ is somewhat
arbitrary, but consistent with the physical picture of our present
Universe entering a de-Sitter phase.
A more reliable estimate of $\eta$ will come from the numerical study
of a model universe with both vacuum and matter energies.
The numerical solution found will be seen to interpolate between
the analytic solutions \eqref{eq:Fsolution} and \eqref{eq:deSsolution1}.

\subsection{Exploratory numerical results}
\label{subsec:Exploratory-numerical-results}

Equation \eqref{eq:4ODEsFRWdim-h} for the potential $u(\dimlessscalar,f)$
from \eqref{eq:dimensionless-potential-u} makes clear that a
model universe with an asymptotically nonvanishing Hubble constant,
$h(\dimlesstime) \to \text{const} \ne 0$,
requires a nonvanishing modified-gravity parameter, $\eta \ne 0$.
The analytic de-Sitter solution with $\dot{h}= \dot{s}=\dot{f} = 0$
has already been given in Sec.~\ref{subsec:Analytic-results}.

\begin{figure*}[ht]
\vspace*{1mm}
\begin{center}  
\includegraphics[width=0.85\textwidth]{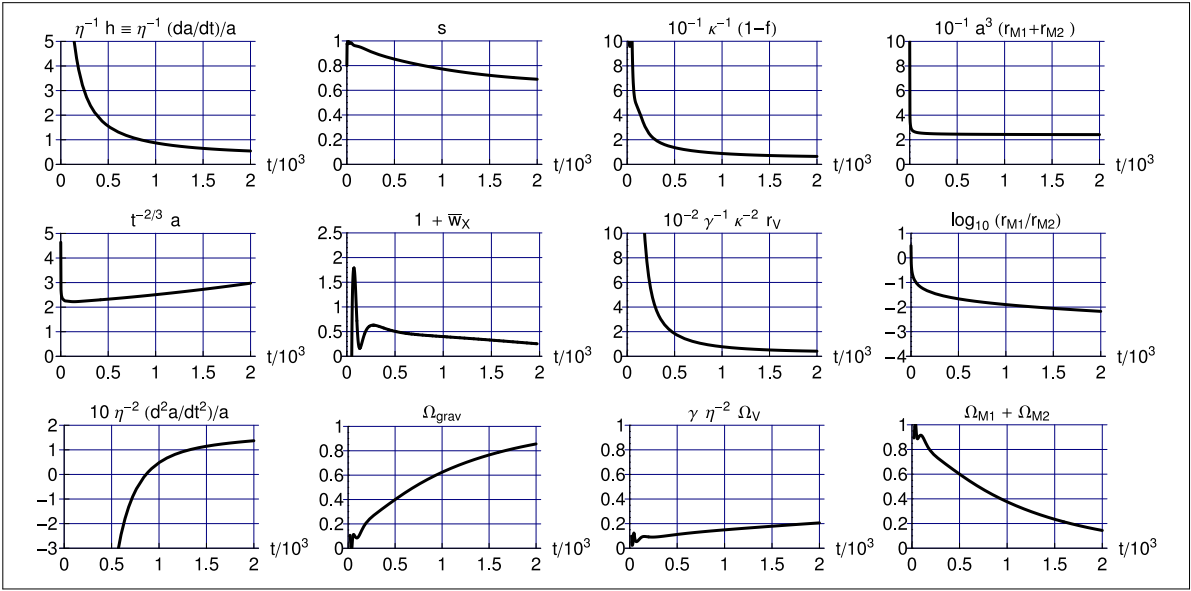}
\end{center}
\vspace*{1mm}
\caption{Numerical solution of ODEs~\eqref{eq:4ODEsFRWdim},
with vacuum energy density \eqref{eq:rV-Ansatz},
Brans--Dicke scalar potential \eqref{eq:dimensionless-potential-u},
and both relativistic matter (energy density $r_{M,1}$) and
nonrelativistic matter (energy density $r_{M,2}$).
The figure panels are organized as follows:
the panels of the first column from the left concern the expansion factor $a(t)$,
those of the second column the modified-gravity scalar $s(t)$,
those of the third column  the gluon-condensate vacuum variable $f(t)$,
and those of the fourth column the matter energy densities $r_{M,n}$.
The model parameters are
$\big(\gamma,\,\eta^2,\, w_{M,1},\,w_{M,2}\big)$ $=$
$\big(                                  {10^{2},\,  9 \times 10^{-7},\,  1/3,\,               0}\big)$,  
with the resulting parameter $\kappa\equiv (3/32)\,\eta^2/\gamma= 8.4375\times 10^{-10}$.
The boundary conditions at $\dimlesstime_\text{start}=0.1$ are
$\big(a,\, h,\, \dimlessscalar,\,  v,\,  1-f,\, r_{M,1},\, r_{M,2}\big)$
$=$
$\big( 1,\, 4.082483,\, 0.8,\, 0.8164966,\, 8.437500\times 10^{-9},\,75.97469,\, 24.02531 \big)$;
see Sec.~\ref{subsec:Elementary-scaling-analysis} for details.
The several energy-density parameters $\Omega$ and
the effective ``dark-energy'' equation-of-state parameter $\overline{w}_{X}$
are defined in \eqref{eq:Omegabar} and \eqref{eq:modgrav-wXbar}, respectively.
With $\gamma/\eta^2 \gg 1$, the values of $\Omega_{V}$ are negligible
compared to those of $\Omega_\text{grav}$ for the time interval shown.}
\label{fig:1}
\end{figure*}

The numerical solution of ODEs~\eqref{eq:4ODEsFRWdim}
for $\eta \sim  10^{-3}$ is presented in Fig.~\ref{fig:1}
and several observations can be made:
\begin{itemize}
\item[(i)]
The boundary conditions on the functions
will be discussed in Sec.~\ref{subsec:Elementary-scaling-analysis}.
\item[(ii)]
There is a transition from deceleration in the early
universe to acceleration in the late universe.
\item[(iii)]
The values for $s$, $1-f$, and $h$ at the largest time shown in Fig.~\ref{fig:1}
agree already at the $10\,\%$
level with those of the analytic de-Sitter-type solution \eqref{eq:deSsolution1}.
\item[(iv)]
The ratio $r_{M,\text{tot}}/\big(6\,h^2\big)$ is equal to $0.25$
at the dimensionless cosmic time $\dimlesstime \approx 1.4\times 10^3$.
\end{itemize}
Points (ii)--(iv) suggest that, for the model parameter values chosen,
the model universe at
$\dimlesstime_{p}= 1.432\times 10^3$ resembles our own present Universe,
characterized by the values \eqref{eq:FRW-OmegaXM-wX}.

More quantitatively, the following three estimates can be obtained.
First, the product of the dimensionless age $\dimlesstime_{p}$
of the present universe with its dimensionless
expansion rate $h(\dimlesstime_{p}) \approx 0.6351\times 10^{-3}$ gives
\bsubeqs\label{eq:results-age-wX-z_inflect}
\beq
t_{p}\,h(t_{p})\approx 0.91\,,
\label{eq:results-age}
\eeq
which also holds for the product of the dimensionful
quantities, $\tau_{p}\,H(\tau_{p})\approx 0.91$.

Second, evaluating the particular combination \eqref{eq:modgrav-wXbar}
of first and second derivatives of $a(t)$ and the matter energy density
$\rho_{M}$, the present effective EOS parameter of the unknown component
is found to be
\beqa
\overline{w}_{X}(\dimlesstime_{p})
&\equiv&
-\frac{2}{3}\,\left.\left(\frac{\ddot{a}\,a}{(\dot{a})^2}+\frac{1}{2}\right)\;
 \frac{1}{1-\Omega_{M}}\;\right|_{\dimlesstime=\dimlesstime_{p}}
          \approx 
-0.66\,.
\label{eq:results-wX}
\eeqa
For larger times $t \gg \dimlesstime_{p}$, this parameter
$\overline{w}_{X}(t)$ drops to the value $-1$, as can be expected from
the right-hand side of \eqref{eq:modgrav-wXbar}.
Additional numerical values are
$\overline{w}_{X}=-0.75082$, $-0.98921$, $-0.99780$, and $-0.99989$
for                 $t=2000$,     $4000$,     $8000$,     and $16000$,
respectively.
Observe that the particular combination of observables \eqref{eq:modgrav-wXbar}
is designed to be interpreted as the effective EOS parameter
of the unknown component $X$ only if matter-pressure effects
are negligible ($t \gtrsim 500$ in Fig.~\ref{fig:1}).

Third, consider the transition of deceleration to acceleration
mentioned in point (ii) above. In mathematical terms, this time corresponds
to the nonstationary inflection point of the function $a(t)$, that is,
the value $t_\text{inflect}$ at which the second derivative of $a(t)$ vanishes
but not the first derivative. Referring to the model universe at
$\dimlesstime_{p}=1.432\times 10^3$, 
the inflection point
$\dimlesstime_\text{inflect}\approx 0.863\times 10^3$  
corresponds to a redshift
\beq
z_\text{inflect}
\equiv
a(t_{p})/a(t_\text{inflect})-1
\approx
0.5\,,
\label{eq:results-z_inflect}
\eeq
\esubeqs
which implies that the acceleration is a relatively recent phenomenon in this
model universe. Inspection of the lower panels of Fig.~\ref{fig:1} shows
that the acceleration sets in when the ratio of
$\Omega_{X}=\Omega_\text{grav}+\Omega_{V}$
and $\Omega_{M,\text{tot}}$ is approximately unity,
whereas the standard $\Lambda$CDM model
would have $\Omega_{X}/\Omega_{M,\text{tot}}\sim 1/2$
according to \eqref{eq:standard-FRW-ddota}.

Returning to the first estimate \eqref{eq:results-age},
note that this quantity can be interpreted as the age of the present universe
in time units obtained from the present expansion rate.
But it is also possible to obtain the absolute age of the model universe,
using the time scale contained in \eqref{eq:Dimensionless1-tau-h}, which
requires as input the experimental value of the QCD gluon condensate $q_{0}$
and the one of Newton's constant $G_{N}$, taken to be equal to
the effective gravitational coupling $\overline{G}_{N}$ from \eqref{eq:Geff}.
With the conversion factors from \eqref{eq:Dimensionless1-tau-h} and the
relation $G \sim s(t_{p})\, G_{N}$
for $K\equiv 1/(16\pi G)$, the numerical results $t_{p} \approx 1432$,
$h(t_{p}) \approx 1/1575$, and $s(t_{p}) \approx 0.7267$
give the following two dimensionful quantities of the present universe:
\bsubeqs\label{eq:results-tp-Hp}
\beqa
\tau_{p} &=& t_{p}\,K\,q_{0}^{-3/4} \sim 13.1\;\text{Gyr}\,,
\label{eq:results-tp}\\[2mm]
H_{p}    &=& h(t_{p})\,K^{-1}\,q_{0}^{3/4}
               \sim 68 \;\text{km}\;\text{s}^{-1}\;\text{Mpc}^{-1}\,,
\label{eq:results-Hp}
\eeqa
\esubeqs
where the numerical values have been calculated
with $q_{0} = (210\;\text{MeV})^4$. Remark that,
if the relation $G \sim G_{N}$ holds for Cavendish-type experiments
as mentioned in \cite{endnote-G_Newton}, the same numerical values
are obtained in \eqref{eq:results-tp-Hp} by taking
$q_{0} \approx (190\;\text{MeV})^4$ and, if $G \sim G_{N}/2 $ holds,
by taking $q_{0} \approx (230\;\text{MeV})^4$.
All of these three $q_{0}$ values lie below the value
$q_{0}\approx (330\;\text{MeV})^4$ indicated by particle 
physics~\cite{ShifmanVainshteinZakharov1978}, but
the uncertainty in the latter value appears to be
large~\cite{Narison1996,Rakow2006,AndreevZakharov2007}.
In addition, it may be that certain particle-physics experiments
are more appropriate than others to
determine the truly homogeneous condensate $q_{0}$ relevant to cosmology.

Next to the
observations~\cite{Riess-etal1998,Perlmutter-etal1998,Freedman2001,Eisenstein2005,Astier2006,Riess2007,Komatsu2008,Vikhlinin-etal2008},
the values obtained in \eqref{eq:results-age-wX-z_inflect}
and \eqref{eq:results-tp-Hp} have the correct order of magnitude,
which is all that can be hoped for at the present stage.
Still, it is remarkable  that more or less reasonable values appear at
all~\cite{endnote-variable-G-Newton}.

For comparison, the standard flat--$\Lambda$CDM
model \eqref{eq:standard-FRW-dota-ddota}--\eqref{eq:FRW-OmegaXM-wX}
with boundary condition $r_{M}(t_{p})/r_{V}=1/3$ gives
the product $\tau_{p}\,H(\tau_{p})\approx 1.01$,
the effective EOS parameter
$\overline{w}_{X}=-1$, and the inflection-point redshift
$z_\text{inflect}=(6)^{1/3}-1 \approx  0.82$.
These three numbers fit the observational data perfectly well, but the
$\Lambda$CDM model is purely phenomenological and cannot explain,
without further input,\footnote{Taking as
additional input the \emph{measured}
value~\cite{Freedman2001} $ h_{0}\approx 0.70$ of the  Hubble constant
$H_{0} \equiv h_{0}\;100\;\text{km}\;\text{s}^{-1}\;\text{Mpc}^{-1}
= h_{0}\, (9.778 \times 10^{9}\,\text{yr})^{-1}$,
the $\Lambda$CDM-model result $\tau_{0}\,H_{0}\approx 1.01$
gives the dynamic age $\tau_{0} \approx 14.2\;\text{Gyr}$.}  
the absolute age of the Universe as in \eqref{eq:results-tp}
or the absolute vacuum energy density as
will be discussed in Sec.~\ref{sec:Conclusion}.

\subsection{Elementary scaling analysis}
\label{subsec:Elementary-scaling-analysis}

In the previous subsection, the ODEs~\eqref{eq:4ODEsFRWdim}
have been solved numerically for certain parameter values and
boundary conditions at $t=t_\text{start}$, which need to be discussed further.
As explained in Sec.~\ref{subsec:Preliminaries},
$t_\text{start}$ is considered to correspond to a time
just after the QCD crossover has happened.
This implies, in particular, that the starting value $h(t_\text{start})$
for the expansion rate is approximately given by
the value $[(r_{V}+r_{M,\text{tot}})/6]^{-1/2}$
of the corresponding standard FRW universe \eqref{eq:standard-FRW-dota}.
The $f$ value at $t_\text{start}$ follows from \eqref{eq:fsolution-B-zeta}
for the chosen $s$ value (see below) and
the starting value for $v$ is obtained by
solving \eqref{eq:Friedmann-type-eq}, considered as a linear equation in $v$
with all other quantities given.

\begin{table}[t]
\begin{center}
\caption{Numerical results for the ``present epoch''
[defined by $\Omega_{M}(t_{p})=0.25$] in model universes
with different numerical values for the parameters $Z$ and $\eta$,
where the latter parameter controls the modified-gravity term
in the action \eqref{eq:action-S-f}
and the former is defined by \eqref{eq:Z-definition} in terms of the
physical energy scales. Other parameters and boundary conditions are given
by \eqref{eq:scaling-gamma-tstart-rM1start-rM2start},
with constants $\widehat{\gamma}$, $\widehat{t}$, and $\widehat{r}$
set equal to $1$. A further boundary condition is $s(t_\text{start})=0.8\,$;
see Sec.~\ref{subsec:Elementary-scaling-analysis} for details.
The effective equation-of-state parameter $\overline{w}_{X}$
and the inflection-point redshift $z_\text{inflect}$ are defined in
\eqref{eq:results-wX} and \eqref{eq:results-z_inflect}, respectively.
Figure~\ref{fig:1} for $Z=10^{-2}$ illustrates the general behavior
of $h(t)$, $\overline{w}_{X}(t)$, and other physical quantities.
\vspace*{2mm}}
\label{tab-scaling-results}
\renewcommand{\tabcolsep}{1pc}    
\renewcommand{\arraystretch}{1.0} 
\begin{tabular}{cc|cccccc}
\hline\hline
$Z$ & $10^{6}\;\eta^2$ &
$10^{-3}\;t_{p}$  & $10^{4}\;h(t_{p})$ & $s(t_{p})$ &
$t_{p}\,h(t_{p})$ & $\overline{w}_{X}(t_{p})$&$z_\text{inflect}$\\
\hline
$10^{-1\phantom{0}}$ & $0.8$ & $1.522$ & $5.980$ &$0.7272$ & $0.910$ & $-0.669$ & $0.541$\\
$10^{-2\phantom{0}}$ & $0.9$ & $1.432$ & $6.351$ &$0.7267$ & $0.910$ & $-0.662$ & $0.538$\\
$10^{-4\phantom{0}}$ & $0.7$ & $1.629$ & $5.584$ &$0.7259$ & $0.910$ & $-0.663$ & $0.515$\\
$10^{-8\phantom{0}}$ & $0.8$ & $1.523$ & $5.967$ &$0.7255$ & $0.909$ & $-0.660$ & $0.505$\\
$10^{-16}$           & $0.9$ & $1.436$ & $6.330$ &$0.7256$ & $0.909$ & $-0.660$ & $0.506$
\\
\hline\hline
\end{tabular}
\end{center}
\vspace*{2cm}\end{table}

Next, the value of $t_\text{start}$ itself and the corresponding
values for $r_{M,1}$ and $r_{M,2}$ need to be specified.
These values depend on the physical ratio $Z$ defined
by \eqref{eq:Z-definition}.
Following the results for the standard FRW universe, take
\bsubeqs\label{eq:scaling-gamma-tstart-rM1start-rM2start}
\beqa
\gamma                              &=&\widehat{\gamma}\;Z^{-1}\,,
\label{eq:scaling-gamma}\\[2mm]
t_\text{start}                      &=&\widehat{t}\;\sqrt{Z}\,,
\label{eq:scaling-tstart}\\[2mm]
r_{M,1}\big(t_\text{start}\big)&=&\widehat{r}\;Z^{-1}\big/\big(1+Z^{1/4}\big)\,,
\label{eq:scaling-rM1start}\\[2mm]
r_{M,2}\big(t_\text{start}\big)&=&\widehat{r}\;Z^{-3/4}\big/\big(1+Z^{1/4}\big)\,,
\label{eq:scaling-rM2start}
\eeqa
\esubeqs
where the constants $\widehat{\gamma}$, $\widehat{t}$, and $\widehat{r}$
are numbers of order unity [in the present elementary analysis,
they are just set equal to $1$]. With $\widehat{t}=1$
and the particular \emph{Ans\"{a}tze}
\eqref{eq:scaling-rM1start}--\eqref{eq:scaling-rM2start},
there is equality of the relativistic (label $n=1$)
and nonrelativistic (label $n=2$) energy densities
around $t\sim 1$, which is not entirely unrealistic if the present universe
has $t\sim 10^3$.

Finally, the boundary condition value $s(t_\text{start})$ is taken
between $0$ and $1$. The results are, however, rather insensitive to
the precise value of $s(t_\text{start})$; see \cite{endnote-sbcs}
for selected numerical results. The explanation is that, independent of the
precise starting value, $s(t)$ increases rapidly until, at $t\sim 1$,
it bounces back from the $s=1$ ``wall''  and, then,
slowly descends towards the de-Sitter value, with some initial oscillations.

Having specified the boundary conditions of the physical variables,
the optimal model parameter $\eta$ needs to be determined.
The strategy is as follows:
for a given $Z$ value, assume an $\eta$ value, determine $t_{p}$
from the condition $\Omega_{M,\text{tot}}(t_{p})=0.25$,
evaluate the product $t_{p}\,h(t_{p})$,
and, if necessary, return to a new value of $\eta$
in order to get $t_{p}\,h(t_{p})$ closer to
the asymptotic value of approximately $0.909$.

Numerical results are given in Table~\ref{tab-scaling-results}.
Three physical quantities, the relative age of the present universe
$t_{p}\,h(t_{p})$,
the effective EOS parameter $\overline{w}_{X}$,
and the inflection-point redshift $z_\text{inflect}$,
appear to approach  constant values as $Z$ drops to zero.
This nontrivial result suggests that the behavior shown
in Fig.~\ref{fig:1} and the corresponding
estimates \eqref{eq:results-age-wX-z_inflect}--\eqref{eq:results-tp-Hp}
also apply to the physical case with $Z\sim 10^{-38}$ as given
by \eqref{eq:Z-definition}.

\section{Conclusion}
\label{sec:Conclusion}

The bottom-row panels of Fig.~\ref{fig:1}, if at all relevant to
our Universe, suggest that the present accelerated expansion may be due
primarily to the nonanalytic modified-gravity term in the
action \eqref{eq:action-S-f} rather than the direct vacuum
energy density $\rho_{V}(q)$, because $q$ is already very close to its
equilibrium value $q_0$, making $\rho_{V}(q)\sim \rho_{V}(q_{0})=0$.
Referring to the definitions in \eqref{eq:Omegabar}, the second panel of
the bottom row shows
the effective energy-density parameter $\Omega_\text{grav}$
due to the gluon-condensate-induced modification of gravity and the third panel
the energy-density parameter $\Omega_{V}$  from the vacuum energy
density proper [with EOS parameter $w_{V}=-1$], their total giving
$\Omega_{X}$ which equals $1-\Omega_{M}$
for a flat FRW universe. As discussed in Secs.~\ref{subsec:Preliminaries}
and \ref{subsec:Exploratory-numerical-results},
the total unknown `$X$' component has an
effective EOS parameter $\overline{w}_{X}$ which drops
to the value $-1$ as the de-Sitter-type universe is approached.

Remark that, in contrast to the results of, e.g.,
Refs.~\cite{Faulkner-etal2007,Brax-etal2008},
nontrivial dark-energy dynamics has been obtained,
because the effective action \eqref{eq:action-S-f} is
assumed to be valid only on cosmological length scales, not solar-system or
laboratory length scales [see also the discussion in the paragraph of
Sec.~\ref{subsec:Theory} containing Eq.~\eqref{eq:h_ext}]. As it stands,
the effective action \eqref{eq:action-S-f} can be viewed as an efficient
way to describe the main aspects of the late evolution of the Universe,
with only two fundamental energy scales, $E_\text{QCD} \sim 10^{8}\;\text{eV}$
and $E_\text{Planck}\sim 10^{28}\;\text{eV}$,
and a single dimensionless coupling constant, $\eta \sim 10^{-3}$. Moreover,
this effective coupling constant $\eta$ can, in principle,
be calculated from quantum chromodynamics and general relativity,
which may or may not confirm our numerical value of approximately $10^{-3}$;
cf. Refs.~\cite{KlinkhamerVolovik2009a,ThomasUrbanZhitnitsky2009}
and the third remark in the Note Added.

Elaborating on the source of the present acceleration, consider the second term
on the right-hand side of \eqref{eq:BDfield-eqs-Gmunu}, which can be
rewritten as $+(2\phi K)^{-1}\,\big(\rho_\text{V,\,BD}\big)\,g_{\mu\nu}$
for the Brans--Dicke vacuum energy density $\rho_\text{V,\,BD}\equiv -K U$.
The exact de-Sitter-type solution \eqref{eq:deSsolution1} for $\kappa \ll 1$,
together with the conversion factor from \eqref{eq:Dimensionless1-u-s}
and Newton's constant from \eqref{eq:Geff}, then allows for the following estimate:
\begin{eqnarray}
\rho_\text{V,\,BD}\,\Big|^\text{(deS,1)}
&=&     -u\,q_{0}^{3/2}/K\,\Big|^\text{(deS,1)}
= 12\pi\,\eta^2\;q_{0}^{3/2}\,G
\sim
(\pi/8)\,\eta^2\; K_\text{QCD}^3/E_\text{Planck}^2
\nonumber\\[1mm] \hspace*{-0.5cm}
&\sim&
\big( 2 \times 10^{-3}\,\text{eV}\big)^4\;   
\times \left(\frac{\eta}{10^{-3}}\right)^2
       \left(\frac{K_\text{QCD}}{\big(420\,\text{MeV}\big)^2}\right)^{3}\,,
\label{eq:rhoV-BD}
\end{eqnarray}
where $q_{0}$ has been expressed in terms of the QCD string tension
$K_\text{QCD}$~\cite{ChengLi1985},
specifically, $q_{0}=E_\text{QCD}^4  \approx (K_\text{QCD}/4)^2$.
The parametric dependence of the above expression,
$\rho_{V}\propto K_\text{QCD}^3/E_\text{Planck}^2$,
is the same as that of the previous estimate (6.7)
in Ref.~\cite{KlinkhamerVolovik2009a}, but expression \eqref{eq:rhoV-BD}
now comes from the solution of field equations.
Two other dimensionful quantities, the age and expansion rate of the
Universe, have already been given in \eqref{eq:results-tp-Hp}.

Before the asymptotic de-Sitter-type universe with effective
energy density \eqref{eq:rhoV-BD} is reached,
the Brans--Dicke scalar $\phi$ evolves and allows for an effective
EOS parameter $\overline{w}_{X}$ different from $-1$
[the scalar $\phi$ has no direct kinetic term in the action \eqref{eq:BDaction-S},
but the $\phi R$ term does give, by partial integration, an effective
kinetic term for $\phi$, which, in fact, leads to the generalized
Klein--Gordon equation \eqref{eq:BDfield-eqs-Box-eta}].
For the present Universe, the general
lesson may be that the deformation of the QCD gluon condensate $q$
by the spacetime curvature of the expanding Universe can result in an effective
EOS parameter $\overline{w}_{X}$ which evolves with time and, for the
present epoch, can still be somewhat above its asymptotic value of $-1$.
In turn, a possible discovery of a $\overline{w}_{X}$ time dependence
may provide an additional incentive to theoretical investigations of the
physics of the gravitating gluon condensate.
\vspace*{-0\baselineskip}\newline
\emph{Note Added. ---} After completion of the work reported here,
we became aware of
two earlier articles and a third article recently posted on the archive.
The first article~\cite{Amendola-etal2007} is a systematic study
of the cosmology of $f(R)$ modified-gravity models and
identifies the modified-gravity term \eqref{eq:action-f},
for constant $q$, as cosmologically viable
[observe the different sign definition of $R$ compared to ours].
The second article~\cite{PogosianSilvestri2007}
investigates the growth of density perturbations
in $f(R)$ modified-gravity models and establishes, in Eq.~(42),
the effective gravitational coupling parameter for
subhorizon CDM density perturbations, which turns out to be
close to $G_{N}$ for the model universe of Fig.~\ref{fig:1}
at times $t \lesssim 500$ (redshifts $z \gtrsim 1$).
The third article~\cite{UrbanZhitnitsky2009} presents a QCD calculation
for the origin of the modified-gravity term \eqref{eq:action-f}
and may also explain the smallness of the coupling constant $\eta$,
even though many conceptual and technical issues remain to be resolved.



\begin{thebibliography}{99}
\bibitem{endnote-sbcs}
Three present-universe quantities have been given
in \eqref{eq:results-age-wX-z_inflect} for $s(t_\text{start})=0.80$,
model parameters $(Z,\,\eta^2)$ $=$ $(10^{-2},\, 9 \times 10^{-7})$,
and further values given by \eqref{eq:scaling-gamma-tstart-rM1start-rM2start}
with constants $\widehat{\gamma}$, $\widehat{t}$, and $\widehat{r}$ set equal to $1$.
For $s(t_\text{start})$ ranging over the interval $[0,\,0.99]$ and all
other inputs kept the same,
the values of $t_{p}h(t_{p})$ vary by approximately $1\,\%$
around the central value,
those of $\overline{w}_{X}(t_{p})$ by approximately $5\,\%$,
and those of $z_\text{inflect}$ by approximately $25\,\%$.
These results suggest that, in the framework of the modified-gravity
theory \eqref{eq:action-S-f}, especially the quantity $z_\text{inflect}$ can
be used as a diagnostic of the state of the Universe after the QCD crossover.
\end{thebibliography}
\end{document}